\documentstyle[12pt]{article}
\input epsf
\begin{document}
\rightline{\vbox{\baselineskip12pt\hbox{EFI-95-61,
USC-95/025}\hbox{hep-th/9509161}}}
\vskip 1cm
\centerline{{\bf \Large \bf
Integrable systems and supersymmetric gauge theory}}

\vskip 1cm
\centerline{{\bf \Large Emil J. Martinec}}
\vskip .4cm
\centerline{\it Enrico Fermi Institute and Dept. of Physics}
\centerline{\it University of Chicago,
5640 S. Ellis Ave., Chicago, IL 60637 USA}

\vskip .4cm
\centerline{\it and}
\vskip .4cm
\centerline{{\bf \Large Nicholas P. Warner}}
\vskip .4cm
\centerline{\it Physics Department, University of Southern
California}
\centerline{\it University Park,  Los Angeles, CA 90089--0484 USA}


\def\pref#1{(\ref{#1})}

\def\ie{{i.e.}}
\def\eg{{e.g.}}
\def\cf{{c.f.}}
\def\etal{{et.al.}}
\def\etc{{etc.}}

\def\inbar{\,\vrule height1.5ex width.4pt depth0pt}
\def\IC{\relax\hbox{$\inbar\kern-.3em{\rm C}$}}
\def\IR{\relax{\rm I\kern-.18em R}}
\def\IP{\relax{\rm I\kern-.18em P}}
\def\Z{{\bf Z}}
\def\A{{\bf A}}
\def\B{{\bf B}}
\def\Pone{{\bf P^1}}
\def\gg{{\bf g}}
\def\hh{{\bf h}}
\def\ee{{\bf e}}
\def\aa{{\bf v}}
\def\h{{h_\gg}}
\def\hdual{{h_\gg^\vee}}
\def\jac{{\sl Jac}}
\def\One{{1\hskip -3pt {\rm l}}}
\def\sutwo{{$SU(2)$}}
\def\nth{$n^{\rm th}$}

\def\xtilde{{\tilde X}}
\def\pic{{\sl Pic}}
\def\hom{{\sl Hom}}
\def\prym{{\sl Prym}}

\def\beq{\begin{equation}}
\def\eeq{\end{equation}}

\def\sst{\scriptscriptstyle}
\def\tst#1{{\textstyle #1}}
\def\frac#1#2{{#1\over#2}}
\def\coeff#1#2{{\textstyle{#1\over #2}}}
\def\half{\frac12}
\def\hf{{\textstyle\half}}
\def\ket#1{|#1\rangle}
\def\bra#1{\langle#1|}
\def\vev#1{\langle#1\rangle}
\def\d{\partial}

\def\np{{\it Nucl. Phys. }}
\def\pl{{\it Phys. Lett. }}
\def\pr{{\it Phys. Rev. }}
\def\ap{{\it Ann. Phys., NY }}
\def\prl{{\it Phys. Rev. Lett. }}
\def\mpl{{\it Mod. Phys. Lett. }}
\def\cmp{{\it Comm. Math. Phys. }}
\def\grg{{\it Gen. Rel. and Grav. }}
\def\cqg{{\it Class. Quant. Grav. }}
\def\ijmp{{\it Int. J. Mod. Phys. }}
\def\jmp{{\it J. Math. Phys. }}
\def\nextline{\hfil\break}
\catcode`\@=11
\def\slash#1{\mathord{\mathpalette\c@ncel{#1}}}
\overfullrule=0pt
\def\AA{{\cal A}}
\def\BB{{\cal B}}
\def\CC{{\cal C}}
\def\DD{{\cal D}}
\def\EE{{\cal E}}
\def\FF{{\cal F}}
\def\GG{{\cal G}}
\def\HH{{\cal H}}
\def\II{{\cal I}}
\def\JJ{{\cal J}}
\def\KK{{\cal K}}
\def\LL{{\cal L}}
\def\MM{{\cal M}}
\def\NN{{\cal N}}
\def\OO{{\cal O}}
\def\PP{{\cal P}}
\def\QQ{{\cal Q}}
\def\RR{{\cal R}}
\def\SS{{\cal S}}
\def\TT{{\cal T}}
\def\UU{{\cal U}}
\def\VV{{\cal V}}
\def\WW{{\cal W}}
\def\XX{{\cal X}}
\def\YY{{\cal Y}}
\def\ZZ{{\cal Z}}
\def\lam{\lambda}
\def\eps{\epsilon}
\def\vareps{\varepsilon}
\def\underrel#1\over#2{\mathrel{\mathop{\kern\z@#1}\limits_{#2}}}
\def\lapprox{{\underrel{\scriptstyle<}\over\sim}}
\def\lessapprox{{\buildrel{<}\over{\scriptstyle\sim}}}
\catcode`\@=12

\begin{abstract}
After the work of Seiberg and Witten, it has been seen that
the dynamics of N=2 Yang-Mills theory is governed by a Riemann
surface $\Sigma$.
In particular, the integral of a special differential
$\lambda_{SW}$ over (a subset of) the periods of $\Sigma$
gives the mass formula for BPS-saturated states.
We show that, for each simple group $G$, the Riemann surface is
a spectral curve of the periodic Toda lattice for the dual group,
$G^\vee$,
whose affine Dynkin diagram is the dual of that of $G$.  This curve
is not unique, rather it depends on the choice of a representation
$\rho$ of
$G^\vee$; however, different choices of $\rho$ lead to equivalent
constructions.  The Seiberg-Witten differential $\lambda_{SW}$ is
naturally expressed in Toda variables, and the N=2 Yang-Mills
pre-potential
is the free energy of a topological field theory defined
by the data $\Sigma_{\gg,\rho}$ and $\lambda_{SW}$.
\end{abstract}

\vskip .5cm
\centerline{{\it Dedicated to the memory of Claude Itzykson,
recalling his gift}}
\centerline{{\it for combining elegant physics and beautiful
mathematics}}

\section{Introduction}

By now, there is a rich `phenomenology' of strong-weak coupling
duality constructions in supersymmetric Yang-Mills theory.
Pure N=2 Yang-Mills theory has been analyzed,
following the seminal work of Seiberg and Witten \cite{seib-wit},
for the groups $A_\ell$ \cite{lerche-yank,arg-far},
$B_\ell$ \cite{swedes},
and $D_\ell$ \cite{germans}.  For each group, one finds a Riemann
surface $\Sigma$, together with a preferred differential
$\lam_{SW}$, such that the spectrum of BPS-saturated states of
electric and magnetic charges $(\vec q,\vec g)$ satisfies
\begin{eqnarray}
  M(\vec q,\vec g)&=&\sqrt2~|\vec q\cdot \vec a +
		\vec g\cdot{\vec a}_D|
	\nonumber\\
  a_i&=&\oint_{{ A_i}}\lam_{SW} \\
  a_{D,i}&=&\oint_{{ B_i}}\lam_{SW} \quad,
	\qquad i=1,\ldots,\rm rank(\gg)\ .\nonumber
\end{eqnarray}
The moduli of the Riemann surface are Casimirs of the gauge group
constructed out of the expectation value of the Higgs field $\phi$.
Its degenerations, and the monodromy of $\vec a$, ${\vec a}_D$
yield a great deal of information, enabling one to reconstruct the
effective gauge pre-potential, $\FF$.  The construction of the
surface and preferred differential has been a somewhat
{\it ad hoc} procedure, and one would like to find a uniform
presentation for all gauge groups.  We will outline such a
construction, in which the differential, $\lam_{SW}$, and the
proper period integrals can be obtained in a canonical and
universal manner.  The complete details will
be postponed for another occasion.

It was noticed independently by one of the present authors,
and by Gorsky \etal\ \cite{russians}, that the $A_\ell$ curve appears as
the spectral curve of an integrable system --
the periodic Toda lattice.  The latter group
went further, pointing out that the Seiberg-Witten differential
appears naturally in the so-called {\it Whitham dynamics} of
adiabatic perturbations of integrable systems,
and analyzing the \sutwo\ case in detail.
We will show that, with a few twists, this framework of Whitham
dynamics for the Toda lattice generates the effective
pre-potential for all gauge groups.

\section{The Toda lattice: the curves}

The theory of integrable systems has become a rich arena for the
interplay of dynamical systems, (loop) group theory,
and algebraic geometry.
The existence of a complete set of integrals of the motion
guarantees the existence of a canonical transformation to
action-angle variables in which time development is straight-line
motion on a torus.  In the Lax formulation,
one realizes the flows generated by these integrals as
\beq
  \frac{\partial \A}{\partial t_i}=[\B_i,\A]\ ,
\label{laxeqs}
\eeq
with the dynamical variables appearing as parameters
in the Lax operator $\A$.
Integrability is a consequence of regarding \pref{laxeqs} as the
condition for a flat gauge potential, so that $\A$, $\B$ are elements
of the Lie algebra $\gg$ of a Lie group $G$.
Group theoretically,
the phase space of the system is a coadjoint orbit of $G$, typically
reduced under the action of some subgroup (frequently Borel).
Particularly interesting cases arise when $G$ is a loop group; then
$\A$, $\B$ are Laurent polynomials in the loop variable $z\in\Pone$.
Given a representation $\rho$, the
Riemann surface $\Sigma_{\gg,\rho}$ defined by the characteristic
polynomial
\beq
  \det[\rho(\A(z))-x\cdot\One]=0
\label{curve}
\eeq
is invariant under the flows.  The torus promised by complete
integrability
is a subspace of the Jacobian $\jac(\Sigma_{\gg,\rho})$ \cite{krichever}.

The specific integrable system whose invariant curve(s) arises
in N=2 Yang-Mills theory is the periodic Toda lattice.
Each affine Dynkin diagram \cite{kac}
defines a periodic Toda lattice \cite{bogo,AvM-painleve}
via Lax operators
\begin{eqnarray}
  \A&=&\sum_{i=1}^{\ell}\left[ b_i \hh_i +
	a_i \ee_{\alpha_i} + \ee_{-\alpha_i}\right]+
	z\ee_{\alpha_0}+z^{-1}a_0\ee_{-\alpha_0}\nonumber\\
  \B_k&=&{\rm Tr}_{\sst L} (\A^k \IP)\ .
\label{todalax}
\end{eqnarray}
The notation is as follows: $z$ is the loop variable;
${\rm Tr}_{\sst L}$ denotes the trace on the left factor of the
$R$-matrix \cite{olive-turok}
\beq
  \IP=\sum_{\alpha\in\;\left\{ {\rm affine~root\atop system}
	\right\}}{\rm sign}(\alpha)\;
	\ee_\alpha\otimes\ee_{-\alpha}\ .
\eeq
The phase space variables are $a_i$, $b_i$; a certain product of
the $a_i$ is left invariant under the flow,
so that the true dimension of
the phase space is $2\ell=2({\rm rank}(\gg)$).
The $\ee_{\alpha_i}$,
$i=1,...,\ell$, are the simple root generators of $\gg$;
the $\hh_i$ span the Cartan subalgebra.
The affine root generator $\ee_{\alpha_0}$ in \pref{todalax}
is the highest (long) root for the usual untwisted Kac-Moody
algebras
$A_\ell^{(1)}$, $B_\ell^{(1)}$, $C_\ell^{(1)}$, $D_\ell^{(1)}$,
$E_{6,7,8}^{(1)}$, $F_4^{(1)}$, $G_2^{(1)}$, given by
\beq
  \alpha_0=\sum_{i=1}^\ell n_i\alpha_i
\eeq
with the integer coefficients $n_i$ for each simple root shown
on the corresponding Dynkin diagram in Table 1.
We will introduce the Toda lattices for
twisted Kac-Moody algebras in a moment.

{\vbox{{\epsfxsize=6.5in
        \nopagebreak[3]
    \centerline{\epsfbox{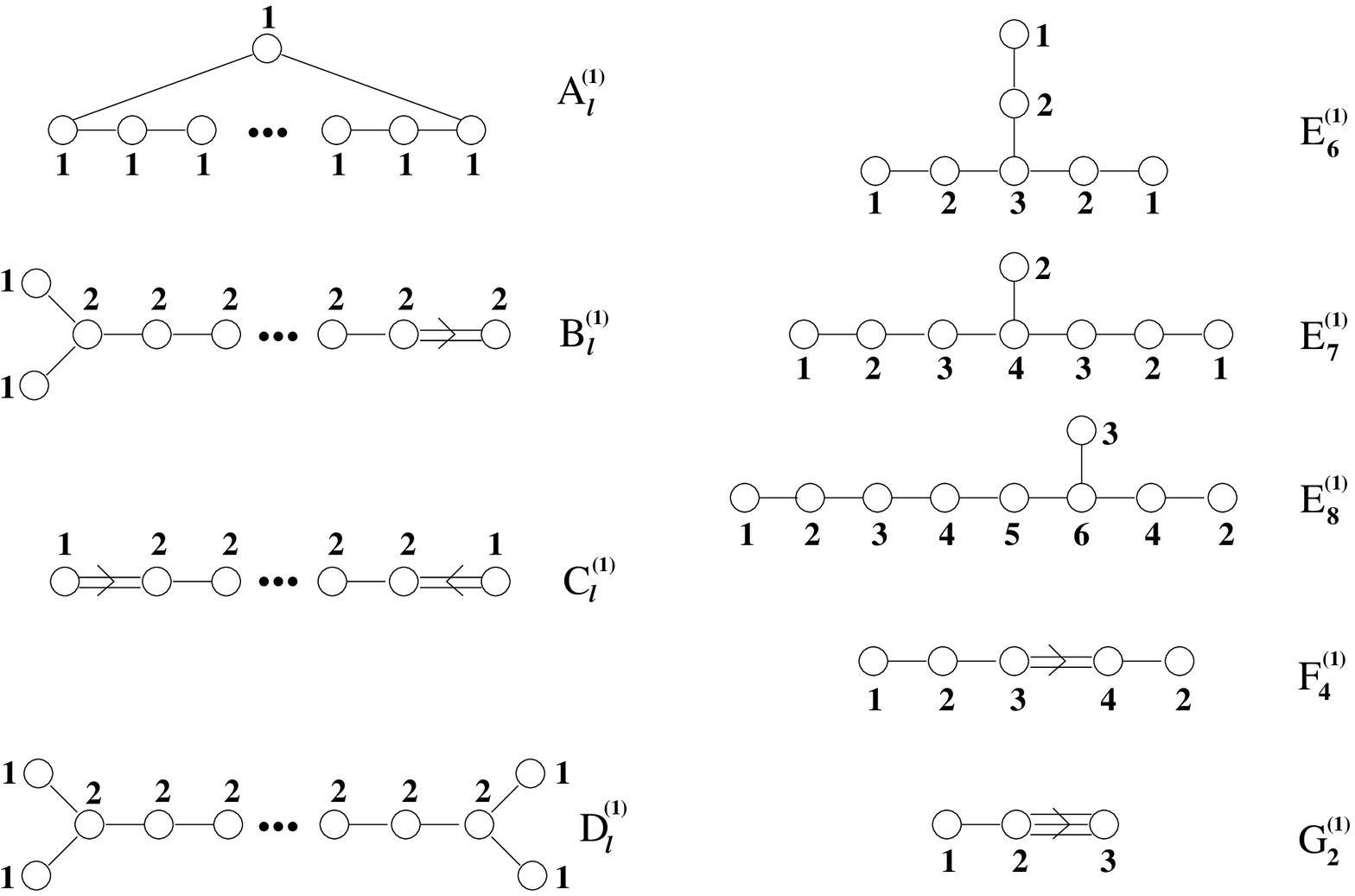}}
        \nopagebreak[3]
    {\raggedright\it \vbox{
{\bf Table 1.}
{\it Dynkin diagrams for untwisted Kac-Moody algebras.  The integers
attached to each node are the Dynkin labels of the affine root.}
 }}}}
    \bigskip}

One might wonder how much of this data used in the linearization of
the Hamiltonian flows is invariantly specified.  In particular, the
construction of the spectral curve $\Sigma_{\gg,\rho}$
requires the choice of a representation $\rho$
that appears neither in the original Hamiltonian equations of motion,
nor in the coadjoint orbit description of the phase space.
Indeed, it is known
\cite{AvM,mcdaniel,mcdaniel-smolinsky,kanev,donagi}
that different choices of $\rho$ lead to different but equivalent
descriptions of the dynamics, via an {\it algebraic correspondence}
(roughly, a one-to-many map in both directions)
between any given pair of choices. This fact will be important to us
below, for instance  one can use it to relate
the curves for $SU(4)$ and $SO(6)$ Yang-Mills that have appeared
in the literature \cite{lerche-yank,arg-far,swedes}.

One of the special things about the Toda lattice is that its
coadjoint orbit is of minimal dimension $2\times{\rm rank}(\gg)$;
thus the constants of motion
are in one-to-one correspondence with the Casimirs of $\gg$.
The root system of $\gg$ respects a grading determined by the
eigenvalue of the Coxeter element of the Weyl group of $\gg$,
or equivalently by the grading determined by the principal
\sutwo\ subalgebra whose Cartan subalgebra generator is
$\delta \cdot \hh$, where $\delta$ is the Weyl vector
($\delta = {1 \over 2} \sum_{\alpha>0} \alpha$).
The equation \pref{curve} will respect this grading; giving weight
one
to $b_i$ and two to $a_i$, the coefficient of $x^k$ in \pref{curve}
must be an invariant of the Lie algebra of the appropriate grade,
hence
is a polynomial in the fundamental Casimirs of $\gg$.
The powers of $z^{-1}$ will occur at grade equal to twice the Coxeter
number
$\h$, multiplied by the trivial invariant $\mu=\prod a_i^{n_i}$
alluded to above; in the Yang-Mills theory, this parameter
sets the quantum scale.  It is also useful to note that
by conjugating $A(z)$, one can find a matrix $\tilde A(z)$ that
satisfies $(\tilde A(z))^t  = \tilde A(1/z)$.  From this
it follows that \pref{curve} is invariant under $z \to \mu/z$.

The list of these curves for the classical
groups with Lax operators in the fundamental representation is:
\begin{eqnarray}
  A_\ell:\quad & z+\mu/z+
x^{\ell+1}+u_2x^{\ell-1}+\ldots+u_{\ell+1}&=0
	\nonumber\\
  B_\ell:\quad & x(z+\mu/z+
x^{2\ell}+u_2x^{2(\ell-1)}+\ldots+u_{2\ell})&=0
	\label{untwisted-KM}\\
  C_\ell:\quad & z+\mu/z + x^{2\ell}+u_2
x^{2(\ell-1)}+\ldots+u_{2\ell}&=0
	\nonumber\\
  D_\ell:\quad & x^2(z+\mu/z) + x^{2\ell}+u_2
x^{2(\ell-1)}+\ldots+u_{2\ell-2}x^2+u_\ell^{\;2}&=0
	\nonumber
\end{eqnarray}
For $A_\ell$ and $D_\ell$, the substitutions $y=z+P(x)/2$ and
$y=x^2 z +P(x^2)/2$, respectively, convert these curves to the forms
presented in \cite{lerche-yank,arg-far} for $A_\ell$ and
\cite{swedes} for $D_\ell$;
however, we will find that the natural Toda variable $z$ is
more convenient.

We immediately see a problem if we are to relate the invariant curves
of Toda dynamics to those of N=2 Yang-Mills theory: the former have
the affine root term $\mu/z$ appearing at a grade $2\h$, whereas the
latter have the instanton-generated term $\mu/z$ appearing at a grade
$2C_2(\gg)=2\hdual$.

\vskip .2cm
\begin{tabular}{||l||c|c|c|c|c|c|c|c|c||}
\hline
group $\gg$ & $A_\ell$ &  $B_\ell$ &  $C_\ell$ &  $D_\ell$ &
	         $E_6$ &  $E_7$ &  $E_8$ &  $F_4$ &  $G_2$ \\ \hline
$\h$	& $\ell+1$ & $2\ell$ & $2\ell$ & $2\ell-2$ &
		  $12$ & $18$ & $30$ & $12$ & $6$ \\ \hline
$\hdual$ & $\ell+1$ & $2\ell-1$ & $\ell+1$ & $2\ell-2$ &
		  $12$ & $18$ & $30$ & $9$ & $4$  \\ \hline
\end{tabular}
\vskip .2cm

\noindent For the simply-laced groups $A_\ell$,
$D_\ell$, and $E_{6,7,8}$, this distinction is irrelevant;
$\h=\hdual$.  However, for the Toda curves of the nonsimply-laced
groups $B_\ell$, $C_\ell$, $F_4$ and $G_2$, the two differ:
$\h\ne\hdual$.
Fortunately, for each nonsimply-laced affine Dynkin diagram,
there is a dual diagram with long roots $\leftrightarrow$
short roots, and $\h\leftrightarrow\hdual$; these are listed in Table
2.

{\vbox{{\epsfxsize=4.5in
        \nopagebreak[3]
    \centerline{\epsfbox{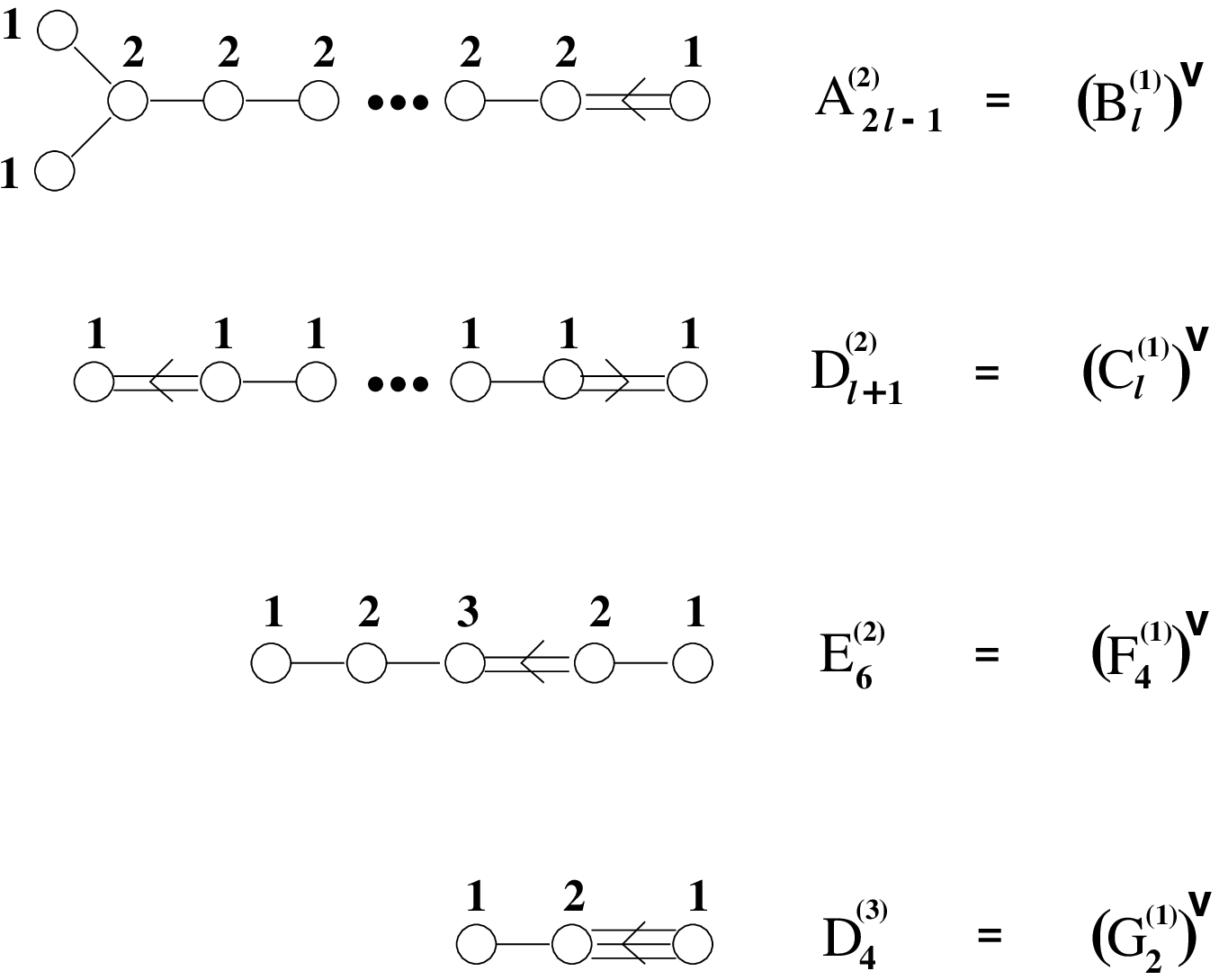}}
        \nopagebreak[3]
    {\raggedright\it \vbox{
{\bf Table 2.}{\it Dynkin diagrams for twisted Kac-Moody
algebras dual to those of Table 1.}
 }}}}
\bigskip}

\noindent The ADE algebras are self-dual under this transformation.
These affine algebras may be obtained as the orbifold
of an untwisted Kac-Moody algebra by the obvious symmetry
of its ordinary Dynkin diagram; thus
$(B_\ell^{(1)})^\vee=A_{2\ell-1}^{(2)}$,
$(C_\ell^{(1)})^\vee=D_{\ell+1}^{(2)}$, $(F_4^{(1)})^\vee=E_6^{(2)}$,
and
$(G_2^{(1)})^\vee=D_4^{(3)}$, where the superscript $(k)$
denotes the order of the diagram automorphism.
The ordinary Lie algebra $\gg_0$ arising from the
untwisted sector is the dual of $\gg$ under the operation
roots $\leftrightarrow$ coroots.  The affine root $\alpha_0$ is the
highest
(short) root in the twisted sector
\beq
  \alpha_0=\sum_{i=1}^\ell m_i \alpha_i\ ,
\eeq
where the coefficients $m_i$ may be read from the table.
With these definitions, the Toda curves for the dual groups are
\begin{eqnarray}
  B_{\ell}^\vee:\quad &
	x(z+\mu/z) +
x^{2\ell}+u_2x^{2(\ell-1)}+\ldots+u_{2\ell}=0\nonumber\\
  C_{\ell}^\vee:\quad &
	(z-\mu/z)^2 +
x^{2\ell}+u_2x^{2(\ell-1)}+\ldots+u_{2\ell}=0\label{twisted-KM}\\
  F_4^\vee:\quad &
	(z+\mu/z)^3+\ldots+P_{27}(x)=0\nonumber\\
  G_2^\vee:\quad &
	3(z-\mu/z)^2-x^8 + 2u_2 x^6 -[u_2^2 +
6(z+\mu/z)]x^4+\nonumber\\
		&\qquad [u_4+2u_2(z+\mu/z)]x^2 = 0\nonumber\ .
\end{eqnarray}
The fact that the twisted Kac-Moody algebra respects a grading
under the `dual' Coxeter element \cite{kac} guarantees that the
`instanton'
terms involving $\mu$ will have grade $[\mu]=2\hdual$.
With the substitution $y=xz+P(x^2)/2$ in the $B_{\ell}^\vee$ curve,
one readily finds the form obtained in \cite{swedes}.

To summarize, to the Montonen-Olive
dual of each simple Lie algebra $\gg$, we may associate
a family of Toda curves $\Sigma_{\gg,\rho}$
whose moduli are the Casimirs of $\gg$, and
which respect the discrete $R$-symmetry of N=2 Yang-Mills.
We will now show how one may pick out a preferred
rank($\gg$)-dimensional
subspace of the Jacobian of $\Sigma_{\gg,\rho}$ on which the
flow linearizes, and the correspondence between them for different
$\rho$.
The associated cycles on $\Sigma_{\gg,\rho}$ will be the special ones
needed for the Seiberg-Witten construction.

\section{Spectral covers, correspondences and the preferred Prym:
the preferred 1-cycles}

To find the spectrum of BPS-saturated states one must compute
$a_i$, $a_{D,i}$ from the periods of a
Seiberg-Witten differential $\lambda_{SW}$ on the Riemann
surface $\Sigma_{\gg,\rho}$.   The problem is
that for all but the fundamental representation of $A_\ell$, the genus
of $\Sigma_{\gg,\rho}$ is much larger than rank($\gg$).  Thus, apart
from finding  $\lambda_{SW}$, one also needs to determine
rank($\gg$) preferred $A$- and $B$-cycles.  This issue has already
been encountered in the study of orthogonal groups, where the
preferred cycles can be isolated using the involution $x \to -x$
\cite{swedes,germans}.  This method generalizes neither to higher
representations nor to the exceptional groups.  However, the general
solution to this problem can be obtained using Toda theories.  The
key is to realize that the coadjoint orbit, and not so much
the surface or its Jacobian, that contains the essential physics.
It is easy to see why this must be
so in the Toda context: There are only $2 \times {\rm rank}(\gg)$
dynamical variables, and since the equations linearize
on the Jacobian, the Hamiltonian flows
select a complex rank($\gg$) dimensional sub-variety of the
Jacobian of any given surface.  This sub-variety turns out to be a
Prym variety, is universal, and in particular is independent of the
representation, $\rho$.  We believe that this Jacobian sub-variety is
the one that solves the Seiberg-Witten problem.  The purpose in this
section is to give a far more usable characterization of this special
Prym variety.

For $SU(N)$, the fundamental representation happens to
give a curve of genus equal to rank($\gg$), and
thus characterizes the special Prym:  it is the
complete Jacobian of this curve.  One can then find how this Jacobian
sits inside the Jacobian of other curves by setting up a
{\it correspondence}.  Given two algebraic curves, $C$
and $\widetilde C$, defined by:
\beq
F(x, z ) ~=~0 \quad {\rm and} \quad \widetilde
F(x^\prime,z^\prime) ~=~0  \ ,
\label{corresp}
\eeq
a correspondence between these curves is a multivalued holomorphic
map given by (one or more) polynomial equations of the form
$\phi_j(x,z,x^\prime,z^\prime)= 0$, $j =1,2, \dots, m$.  A
correspondence can be used to map homology cycles and holomorphic
differentials from one curve to another, and in particular maps the
special Prym of $C$ to the special Prym of $\widetilde C$ --
therefore the fundamental representation of $SU(N)$
generates the special Prym in any representation.
This is discussed extensively in \cite{mcdaniel}, where
the example of $SU(4)$ in the $\underline{\bf 4}$ and in the
$\underline{\bf 6}$ is discussed in detail.  Indeed, suppose that
$F(x,z)=0$ defines the curve for the fundamental of $SU(N)$, and
$\widetilde F(x^\prime,z^\prime)=0$ defines the curve for the two
index anti-symmetric tensor of $SU(N)$.  Then a correspondence
between
these curves is given by
\beq
z ~=~ z^\prime \quad {\rm and} \quad  F(x - x^\prime,z) ~=~0  \ .
\eeq
For $N=4$ this correspondence maps the Jacobian of the fundamental
curve of genus $3$, to a three dimensional sub-variety of the the
Jacobian of the genus $5$ curve derived using the
$\underline{\bf 6}$ of $SU(4)$.  One can check that the corresponding
cycles are precisely those identified in \cite{germans} as the
solution
of the Seiberg-Witten problem for $SO(6)$, and that $\lambda_{SW}$
for one curve maps to $\lambda_{SW}$ for the other.

This method of correspondences is subsumed in a more general group
theoretic procedure.  The first step is to view the Riemann surface
(\ref{curve}) as a foliation over the Riemann sphere for $z$, and
extend $x$ to an analytic function of $z$.  The number of
leaves is then equal to the dimension of the representation $\rho$.
Leaves are connected when $z$ takes a value at which two of the
eigenvalues of $\A(z)$ become the same.
Since the determinant
(\ref{curve}) is invariant under $z \rightarrow \mu/z$,
such values of $z$ occur in pairs related by $z \to \mu/z$.
At a generic Higgs vev $\varphi=\vev{\phi}$, and for
finite $x$, the eigenvalues come together in pairs, and the result is
a square-root branch cut running between the two values of $z$ at
which this happens.  It is now convenient to consider the element
$\aa \cdot \hh$ of the Cartan sub-algebra of $\gg$ that is
conjugate to $\A(z)$.  By using the action of the Weyl group, we will
take $\aa \cdot \hh$ to be in the fundamental Weyl chamber.  If
$w^{(p)}$ are the weights of the representation $\rho$, then the
eigenvalues of $\A(z)$, and hence the leaves of the foliation, are
given by $x = w^{(p)} \cdot \aa$, $a =1, \dots, dim(\rho)$.
Since Casimir invariants are equivalent to Weyl invariant polynomials
in the weights, it follows that in general (\ref{curve}) will
decompose into disconnected component Riemann surfaces associated
with the Weyl orbits of the weights of $\rho$.  So we will focus on
one of these components.  The sheets of the foliation will
generically come together when $\aa \cdot \hh$ ceases
to be a regular element of the Lie algebra,
that is, when it hits a wall of the fundamental Weyl chamber.  At
generic Higgs vevs in the fundamental Weyl chamber,
this happens when $\aa \cdot \alpha_i = 0$,
for some simple root $ \alpha_i$.  There can be
also be ``accidental'' singularities where two eigenvalues come
together, and yet $\aa \cdot \hh$ is regular.  Whether this happens
or not depends upon the details of the specific example, and
even if it happens, there is a method of constructing a cover
of the Riemann surface with accidental singularities such that
the cover only has branch points when $\aa \cdot \hh$ ceases
to be regular \cite{donagi}.  We will therefore ignore
accidental singularities, apart from a brief comment later on.

There are thus rank($\gg$) branch cuts over the $z$-plane,
each associated
with a simple root of $\gg$.  Above the cut corresponding to
$\alpha_i$,
leaves labelled by $w^{(p)}$ and $w^{(q)}$ will be joined together
along the cut if and only if the Weyl reflection generated by
$\alpha_i$ exchanges the the weights $w^{(p)}$ and $w^{(q)}$.
This characterizes the foliation at finite values of $x$.  There is
another cut that runs from $z=0$ to $z=\infty$ which determines
the structure at $x = \infty$.  Taking the limit as $x \to \infty$
in (\ref{curve}), and keeping the dominant terms, one finds that the
result has the Coxeter symmetry of the dual affine Lie algebra,
and that the leaves in the foliation are connected under the
action of this symmetry.

The whole point of this construction is that the resulting surface
has a symmetry action of the Weyl group acting upon it.  This can be
used to induce a symmetry on the cycles, upon the
holomorphic differentials and hence upon the Jacobian variety.
However, even though the leaves in the
foliation are a Weyl orbit,
the Weyl representation can be (and in fact is) reducible;
moreover, it generically contains the fundamental
rank($\gg$) dimensional reflection representation.  The preferred
Prym is obtained by passing to the part of the Jacobian variety that
corresponds to the reflection representation of the Weyl group.
In a moment we will give an example, but
we first wish to make the foregoing statement more precise by
quoting the results of \cite{donagi}.

Consider a principal $G$-bundle $\VV\rightarrow X$ over an
algebraic variety $X$, together with a section $\phi$ of the
adjoint bundle $\VV\times_G\gg$
(\ie, a Higgs vev)\footnote{Note that this
construction is much more general than the consideration of a
spectral curve and integrable flows on its Jacobian; $X$
can be {\it any} algebraic variety, leading to the notion
of an {\it algebraically completely integrable} Hamiltonian system.}.
Given this data, one can construct \cite{kanev,donagi} a kind of
`universal spectral cover' $\xtilde$ with Galois group $W$,
the Weyl group of $G$.
The points of $\xtilde$ over $z\in X$ can typically
be identified with chambers of the dual of the unique Cartan
subalgebra $\hh(\phi(z))$ containing $\phi(z)$.
For instance, when $G$ is $SU(N)$ considered as a matrix
in the fundamental representation, a point of the fiber is
an ordering of the eigenvalues of $\phi(z)$.
The universal cover $\xtilde$ is independent of a choice of
representation.

Given a representation $\rho:G\rightarrow GL(V)$,
with highest weight $\lambda$, we can construct a
surface $\xtilde_\lam$ as outlined above; or we can
construct a quotient space $\xtilde_P=\xtilde/W_P$, where
$W_P\subset W$ is the parabolic subgroup that fixes $\lam$.  If there
are no accidental singularities then $\xtilde_\lam$ is isomorphic
to $\xtilde_P$, otherwise $\xtilde_\lam$
is birational to $\xtilde_P$, with $\xtilde_P$ covering
$\xtilde_\lam$.
Consider the group ring (\ie\ the regular representation) of $W$
and its decomposition into irreps $\Lambda_i$
\beq
  \Z[W]\sim\oplus_i\  \Lambda_i\otimes_\Z \Lambda_i^*\ .
\eeq
Using $W$-equivariant maps from $\Lambda_i$ into $\pic\xtilde$,
$\prym_{\Lambda_i}\equiv\hom_W(\Lambda_i,\pic\xtilde)$,
one obtains decompositions of $\pic\xtilde$ and $\pic\xtilde_P$
\begin{eqnarray}
  \pic\xtilde\sim&\bigoplus_i\
\Lambda_i\otimes_\Z\prym_{\Lambda_i}(\xtilde)
	\nonumber\\
  \pic\xtilde_P\sim&\bigoplus_i\
M_i\otimes\prym_{\Lambda_i}(\xtilde)\ .
\end{eqnarray}
Here $M_i$ is the subspace of $\Lambda_i$ left fixed by $W_P$.
So to any spectral curve $\xtilde_P$, there is a natural
decomposition
of its Prym variety under irreducible representations
of the Weyl group.  In particular, one has \cite{kanev,donagi}:

\begin{itemize}
\item {\it The Prym variety $\prym_\Lambda(\xtilde)$
corresponding to the reflection representation $\Lambda$ (\ie\
the weight space of $G$) occurs with positive multiplicity
in $\pic(\xtilde_P)$ for any proper subgroup $W_P\ne W$.}
\end{itemize}

In other words, we can always find the Cartan torus inside
the Picard variety of the desingularization $\xtilde_P$ of
$\xtilde_\lam$.
Kanev\cite{kanev} and Donagi\cite{donagi} go on to construct a
projection
operator onto this preferred Prym.  Finally, the integrable
flows linearize on $\prym_\Lambda(\xtilde)$ \cite{kanev}.

Consider now the example of $SU(5)$ in the ten dimensional
representation.
We present this example because it displays most of the features
of the generic case (and provides a useful warmup for $E_6$
in the $\underline{\bf 27}$).
The polynomial (\ref{curve}) is quadratic in $(z+\mu/z)$, and
can be factored as
\begin{eqnarray}
  z + \mu/z &=&  u_5 + (p_0 \pm p_1 \sqrt{p_2})/2 \ ;\nonumber\\
p_0 &=& 11 x^5 + 4 u_2 x^3 + 7 u_3 x^2 + (u_2^2 - 4 u_4) x  +
u_2 u_3\ ; \nonumber \\
p_1 &=& 5 x^3 + u_2 x + u_3 \ ;  \nonumber \\
p_2 &=& 5 x^4 + 2 u_2 x^2 + 4 u_3 x + u_2^2 - 4u_4\ .
\label{tencurve}
\end{eqnarray}
One can proceed directly and think of
the surface as a four-fold cover of the $x$-plane.  Counting
the branch points, a simple application of Riemann-Hurwitz
shows that the surface has genus 11.

Following the discussion above, it is more natural to think
of the surface as a ten-sheeted cover over the $z$-plane, with
four branch cuts of order $1$ ({\it i.e.} square root cuts)
at finite $x$, and one branch cut of order $4$
({\it i.e.} $x^5 \sim y$) between $z=0$ and $z=\infty$.
Label the weights of the $\underline{\bf 10}$, and hence the
leaves in the foliation, by vectors of the form $(1,1,0,0,0) +
{\rm permutations}$ (see Figure 1.).
The Coxeter element, $s$, is the cyclic
permutation of order $5$, and collects the weights of the
$\underline{\bf 10}$ into two groups of $5$.  This specifies
the connections between the foliations at the cut between
$z=0$ and $z=\infty$.  For each simple root $\alpha_i$,
$i = 1,\dots,4$, there are exactly three pairs of weights
that are interchanged by the Weyl reflection $r_{\alpha_i}$,
and corresponding leaves are connected in pairs, leading to
$12$ pieces of interconnecting plumbing.  If one takes into
account the connections from the cut $z=0$ and $z=\infty$,
the ten sheets connect into two disjoint two spheres, and the
remaining $12$ connections, associated with the simple roots,
convert this into a genus $11$ surface.  The fact that we get
the same answer -- viewing the surface as a foliation over
the $z$-plane, and as a branched cover of the $x$-plane
using \pref{tencurve} -- indicates that there
are no accidental singularities.  The resulting picture is
shown in Figure 1.

\bigskip
{\vbox{{\epsfxsize=4.5in
        \nopagebreak[3]
    \centerline{\epsfbox{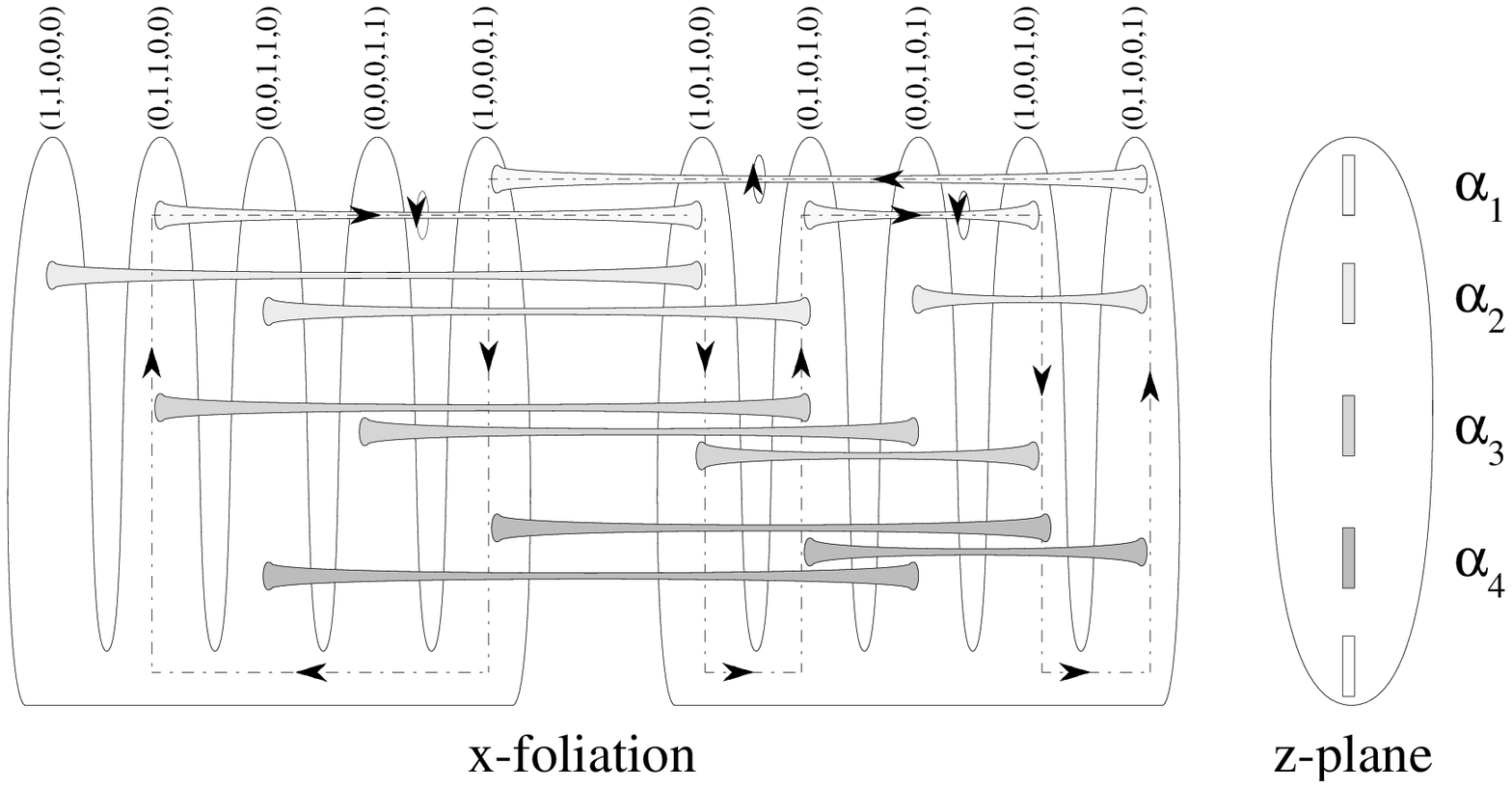}}
        \nopagebreak[3]
    {\raggedright\it \vbox{
{\bf Figure 1.}{\it Spectral curve for $SU(5)$
in the {\bf $\underline{10}$}.  Typical $A$ and $B$ cycles
of the Toda flow (associated to the $\alpha_1$ cut) are indicated.}
 }}}}
    \bigskip }

It is relatively easy to find the reflection representation
of the Weyl group of $SU(5)$ inside the ten-dimensional
representation of $SU(5)$:  one defines $x_i = \sum_\lambda
(\lambda \cdot \alpha_i) v_\lambda$, where $v_\lambda$ is the
basis vector of the $\underline{\bf 10}$ with weight
$\lambda$.  Under the action of the Weyl group on the
$v_\lambda$, the vectors $x_i$ transform as if they were the
vectors $\alpha_i \cdot \hh$ in the
Cartan sub-algebra -- indeed, the inner product matrix of the
$x_i$ is a multiple of the Cartan matrix.  One can easily check
that combining leaves in the foliation in this manner is
very closely related to action of the correspondence mapping
(\ref{corresp}).

There is a natural method for identifying a preferred
set of $A$-cycles.  Define $A_i$ to be lifts of the cycles that
run around the $\alpha_i$-cuts in the $z$-plane.
Specifically, above each such cut in the
$z$-plane there are three connections between leaves, about
each of which we can draw a cycle.  (The orientation of the
cycles in the lift may be fixed by choosing it to be
anti-clockwise in the leaf whose weight has positive inner
product with $\alpha_i$.)
The $A_i$ transform in the reflection representation of the
Weyl group, and thus define the preferred Prym.
The associated $B$-cycles also arise from the $z$-plane;
the obvious guess works --
one lifts a curve that runs around $z=0$ and one of the
$\alpha_i$ branch points.  That is, one can construct a curve
$B_i$ that only  passes through the plumbing at $x = \infty$, and
through the $\alpha_i$ branch cuts (for fixed $i$), in such a
manner that it has an intersection number $+1$ with the three
disjoint pieces of $A_i$.  One can describe this cycle, $B_i$,
group theoretically.  An unreduced product of elements of
the Weyl group (a ``word'') can be viewed as defining a path
by starting on some leaf and then connecting to other leaves
in a manner defined by the basic Weyl reflections that make up
the element.  Consider $s_{\alpha_i} = s r_{\alpha_i}$,
where $r_{\alpha_i}$ is the simple reflection of the root $\alpha_i$
and $s$ is the Coxeter element;
use it to define a path which only makes connections
through the $\alpha_i$ cut, or through infinity.  Since the
element $s_{\alpha_i}$ has order $4$, the word $s_{\alpha_i}^4$
defines a closed loop.   The result is four closed cycles,
$B_i$, whose intersection numbers with the $A_j$ are
$B_i \cdot A_j = 3 \delta_{ij}$.  (It is important to choose
the  proper Coxeter element here.)  While we have only verified
this construction in some examples, we believe that this
will provide an explicit construction of the preferred Prym
in all generality.

\section{Topological field theory: the BPS formula
and the N=2 Yang-Mills pre-potential}

Having constructed a Toda spectral curve with the symmetries
of the N=2 Yang-Mills effective theory, and a special Prym variety
with the structure of Yang-Mills electric and magnetic weights,
the next logical step is to look for natural objects in the
Toda theory which correspond to the Seiberg-Witten differential
$\lam_{SW}$ and the gauge pre-potential.  The answers are quite
striking: First, $\lam_{SW}$ is the action differential `$p\,dq$'
for a natural conjugate pair of Toda variables, so that the integrals
$a_i=\oint_{{\rm A_i}}\lam_{SW}$
are action coordinates for Toda; second, the N=2 gauge pre-potential
is the free energy $\FF=\log(\tau)$
of the (Dubrovin-Krichever) topological field theory
of adiabatic deformations on the moduli space
of Toda dynamics!\footnote{On this last point, it was
I. Krichever who stressed to us the topological field theory
interpretation of \cite{russians} (private communication).}

The transformation of the Toda evolution to action-angle
variables proceeds in two steps.
First one defines an `auxiliary spectrum' -- the zeroes $x_i$ of
a fixed eigenvector $\psi_0$ of the Lax operator $\A$.
The analysis of \cite{flaschka-mclaughlin,vanM,AvM,AHH}
shows that these zeroes travel along the special
cycles of the curve, and hence provide a local system of coordinates
on $\prym_\Lambda$ (via the Abel map).
The canonical conjugate momenta $p_i$ take the values
$\log[z(x_i)]$ on the classical motion.
The solution $z(x)$ to $P(x,z)=0$, Eq. \pref{curve},
defines the spectral curve $\Sigma_{\gg,\rho}$;
this equation may then be interpreted
as a kind of dispersion relation.
%
The orbit invariants (action integrals) are
\beq
  J_i=-\frac{1}{2\pi}\oint_{i^{th}~\rm A-cycle}\log[z(x_i)]\;dx_i\ .
\label{orbit-invt}
\eeq
In the second step,
one derives the angle variables as $\theta_i=\d S/\d x_i$ from the
action
\beq
  S(x,J)=\int^x \log[z(x')]\;dx'\ ,
\eeq
as well as the oscillation frequencies $\omega_i={\dot\theta}_i=\d
H/\d J_i$.
The intermediate variables $x$, $z$ are not merely useful in
solving for the Toda flows; in the Yang-Mills theory
the orbit invariants \pref{orbit-invt} are precisely the
$a_i$ that enter the BPS mass formula!  It is easy to verify that
the formulae for $\lam_{SW}$ in the literature on the particular
groups
$A_\ell$\cite{lerche-yank,arg-far}, $B_\ell$\cite{swedes},
and $D_\ell$\cite{germans}  all take the form
\beq
  \lam_{SW}=x\;d\log(z)\simeq -\log(z)\;dx
\eeq
when expressed in Toda variables.  Hence we may identify
\beq
  \lam_{SW}\equiv dS_{\rm Toda}\ .
\eeq
We should mention that much of the work on these conjugate
variables is specific to $A_\ell$ Toda (the other classical groups
are considered in \cite{AvM}); however we are confident that the
results are general.

The final entry in our Toda/Yang-Mills lexicon is the effective
N=2 pre-potential $\FF(a,a_D)$.  Its characteristic properties are
\begin{eqnarray}
  a_{D,i}&=&\frac{\d \FF}{\d a_i}		\nonumber\\
  \frac{\d a_{D,i}}{\d a_j}&=&\frac{\d^2\FF}{\d a_i\;\d a_j}
	 \quad=\quad \tau_{ij}\ ,
\label{properties}
\end{eqnarray}
where $\tau_{ij}$ is the period matrix of the Riemann surface
$\Sigma_{\gg,\rho}$.  Given an integrable hierarchy,
Krichever\cite{krich-cargese,krich-CPAM} and
Dubrovin\cite{dubrovin-npb}
have constructed a topological field theory having precisely
these properties; this topological field theory describes the
dynamics of the {\it Whitham-averaged} hierarchy.  Starting with
the periodic motion of the original integrable system, one may
ask whether it is possible to adiabatically perturb
the integrals of motion (\ie\ the spectral curve $\Sigma_{\gg,\rho}$)
in a consistent manner -- such that the slow perturbation of the
integrals of motion is `orthogonal' to the fast orbital motion.
In other words, one wants to find a dynamics on the moduli space of
the original system (the procedure thus has strong analogies to
the quantization of collective coordinates, and also to
low energy monopole scattering \cite{atiyah-hitchin}).
This nonlinear version of the WKB method starts with an
eigenfunction of the unperturbed problem
\beq
  \psi_0(t_\alpha)=\psi_0(k_\alpha t_\alpha|u_1,...,u_\ell)\ ,
\eeq
where the $u_i$ are the moduli of the spectral curve (the
constants of the motion);
then one takes an ansatz for the perturbed solution
\beq
  \psi(T_\alpha)=\psi(\eps^{-1}S(T_\alpha)|u_i(T_\alpha))\ .
\eeq
It is useful to keep in mind the usual linear WKB method,
where $exp[i(kx+\omega t)]$ is replaced by $exp[iS(x,t)/\eps]$.
Taking $T_\alpha=\eps t_\alpha$ and $k_\alpha=\d S/\d T_\alpha$,
the ansatz satisfies the equations of motion to lowest
order in $\eps$.  Expanding to next order gives a set of
consistency conditions on the averaged densities
${\overline\Omega}_\alpha$ of the conservation
laws
\beq
  \d_\alpha{\overline \Omega}_\beta=\d_\beta{\overline\Omega}_\alpha\
,
\label{whitham-flows}
\eeq
and implies ${\overline\Omega}_\alpha=\d(dS)/\d T_\alpha$;
\ie, the averaged hierarchy is again an integrable hierarchy.
Dubrovin\cite{dubrovin-npb} and Krichever\cite{krich-CPAM}
show that the tau function for this hierarchy
satisfies the axioms of topological field theory, and has the
properties \pref{properties}.
Thus we may identify \cite{russians}
\beq
  \FF_{\rm N=2~YM}\equiv\log({\tau}_{\sst\rm Toda-Whitham})\ .
\eeq

\section{Discussion}

We have seen that there exists a uniform structure underlying the
effective action of N=2 Yang-Mills theory, involving integrable
systems and topological field theory.  We have not yet
understood how to derive this structure from the underlying
Yang-Mills
dynamics.  The situation is rather reminiscent of the $c< 1$
matter-gravity theories\cite{2dgrav}.
There one has an underlying field theoretic
description involving Liouville field theory coupled to
the matter system (at least near fixed points).  It is very difficult
to compute in this theory.  However, if one
is interested in global questions such as integrated correlations of
order parameters, there exists an elegant solution
in the form of $\tau$-functions, the KP hierarchy, and topological
field theory\cite{2dgrav}.
The connection between this construction and the field-theoretic
formulation is quite indirect.
Answers to `local' (nevertheless gauge invariant)
questions, such as the Ising correlation function at fixed geodesic
distance or the partition function at fixed surface moduli,
are harder to come by; they are not encompassed by the
integrable-topological formalism
and the only tool at present is to resort to the field theoretic
description.
On the other hand, this zero-momentum sector contains a great
deal of information about the phase structure of the theory.
A great deal is known about the singularities of Toda dynamics
\cite{AvM-sings} which might be adapted to the study of
degenerations in N=2 Yang-Mills \cite{arg-doug}.

One may contemplate a much broader applicability of
the above ideas.  Curves and Seiberg-Witten differentials
for \sutwo\ theory with fundamental and
adjoint matter\cite{seib-wit-2}, and for
$SU(N)$ with fundamental matter\cite{hanany-oz,arg-plesser,nem-min},
have been proposed.
They have natural forms when expressed in Toda-style variables.
Due to the discrete $R$-symmetry,
the effect on the spectral curve of adding matter
is to reduce the degree of the $\mu/z$ term from $2C_2(G)=2\hdual$ to
$[2C_2(G)-\sum_i T_2(\rho_i)]$, with the deficit in homogeneity
weight
made up in these examples
by a polynomial $G(x,m_i)$ of degree $\sum_i T_2(\rho_i)$.
The obvious modification of the Toda Lax operator to try is to
lower the grade of the highest weight element $z^{-1}a_0
\ee_{\alpha_0}$.
This will simultaneously
increase the size of the coadjoint orbit, so that the moduli
of the curve may include the $m_i$.

The appearance of Kac-Moody algebras in the solution of N=2 gauge
theories is a profound mystery.
The fact that the dual $G^\vee$ of the gauge group $G$
appears suggests a connection to monopole dynamics.
The appearance of the spectral
parameter (loop coordinate) $z$ begs a physical explanation.
This sort of structure has been encountered in $N=2$ supersymmetric
models in two dimensions, where the dependence of
certain two-point functions upon couplings and scales, and
even two-dimensional instanton corrections, is
determined by some classical integrable equations (the $t - t^*$
equations) \cite{SCCV}.  Following similar lines of
thought, one is naturally led to consider the four-dimensional
version of the chiral ring, and its action on the bundle of
Yang-Mills vacua over the base of Higgs vevs.  One finds many
close parallels, and some significant differences, between the
two and four-dimensional structures, and this approach is
still under investigation.  The fact that there is a duality between
two and four dimensional instanton corrections
in string theory \cite{sqms} strongly suggests
that there should indeed be a four dimensional version of $t-t^*$.
The appearance of affine Lie algebras
also suggests that the level-rank dualities
of WZW models \cite{levelrank}
might `explain' N=1 Yang-Mills dualities.
In topological field theories closely related to ours
\cite{vafa,dubrovin-npb,krich-CPAM}, a useful
set of coordinates localizes the dynamics on the critical
points (branch points) of the curve; this may be the convenient
way to describe soft breaking to N=1 \cite{seib-wit,arg-doug}.
The soft breaking terms
could be added to the Whitham dynamics as a secular
perturbation of the flow equations
\pref{whitham-flows} \cite{krich-cargese}.

One might also imagine extending the scope of our construction
to string theory.  Kachru \etal\ \cite{lerche-vafa} have found the
spectral curve of rigid N=2 Yang-Mills theory as a limit in the
Calabi-Yau
moduli space of some examples, but not others.
The role of the periods of the Seiberg-Witten differential is
played by periods of a three-form on the Calabi-Yau.
This is not inconsistent with our results; the invariant
specification
of Toda as coadjoint orbit dynamics on a loop group
makes no reference to Riemann surfaces.
Notice also that the notions of spectral cover and
analytically completely integrable Hamiltonian system
are quite general,
and not at all restricted to $X$ being a Riemann surface;
it is not out of the question that one could construct an
integrable system based on Calabi-Yau manifolds
that `explains' the results of Kachru \etal\cite{lerche-vafa},
and that includes the effects of coupling to supergravity.
Donagi and Markman \cite{donagi-markman} have built
an integrable system canonically associated to any Calabi-Yau
threefold, and conjecture that it is related to
mirror symmetry (and, one might expect, all of T-duality).
This Calabi-Yau integrable system bears many similarities
to its Riemann surface counterpart employed above.
Indeed, could there not be a grand integrable system that governs
all the miraculous stringy dualities that have emerged?

\vskip 1cm
\noindent {\bf Acknowledgements:} We are grateful to
K. Intriligator for discussions of supersymmetric gauge theory,
and especially I. Krichever for explanations of his work.
E.M. would like to thank U. of Paris VI and CERN,
and both authors thank the Aspen Center for Physics, for their
hospitality during the course of this work.
E.M and N.W. are supported in part by funds provided by the DOE under
grant Nos. DE-FG02-90ER-40560 and DE-FG03-84ER-40168, respectively.

\end{document}